\documentclass[%
 reprint,		
 amsmath,amssymb,superscriptaddress,
 aps,
 prmaterials,
]{revtex4-2}

\usepackage{graphicx}
\usepackage{dcolumn}
\usepackage{bm}
\usepackage{hyperref}
\hypersetup{colorlinks=true, citecolor=blue, urlcolor=blue, linkcolor=blue}
\begin{document}

\preprint{APS/123-QED}

\title{Ternary Hypervalent Silicon Hydrides via Lithium at High Pressure}
\author{Tianxiao Liang}	
\affiliation{College of Physics, Jilin University, Changchun 130012, China}
\author{Zihan Zhang}	
\affiliation{College of Physics, Jilin University, Changchun 130012, China}
\author{Xiaolei Feng}	
 \email{xf232@cam.ac.uk}
\affiliation{Center for High Pressure Science and Technology Advanced Research, Beijing 100094, China}
\affiliation{Department of Earth Science, University of Cambridge, Downing Site, Cambridge CB2 3EQ, United Kingdom}
\author{Haojun Jia}	
\affiliation{Department of Chemistry, Massachusetts Institute of Technology, Cambridge, Massachusetts 02139, USA}
\author{Chris J. Pickard}	
\affiliation{Department of Materials Science \& Metallurgy, University of Cambridge, 27 Charles Babbage Road, Cambridge CB3 0FS, United Kingdom}
\affiliation{Advanced Institute for Materials Research, Tohoku University 2-1-1 Katahira, Aoba, Sendai, 980-8577, Japan}
\author{Simon A. T. Redfern}	
\affiliation{Asian School of the Environment, Nanyang Technological University, Singapore 639798}
\author{Defang Duan}	
 \email{duandefang@jlu.edu.cn}
\affiliation{College of Physics, Jilin University, Changchun 130012, China}

\begin{abstract}
Hydrogen is rarely observed as ligand in hypervalent species, however, we find that high-pressure hydrogenation may stabilise hypervalent hydrogen-rich materials. Focussing on ternary silicon hydrides via lithium doping, we find anions composed of hypervalent silicon with H ligands formed under high pressure. Our results reveal two new hypervalent anions: layered-SiH$_{5}^{-}$ and tricapped trigonal prismatic SiH$_{6}^{2-}$. These differ from octahedral SiH$_{6}^{2-}$ described in earlier studies. In addition, there are further hydrogen-rich structures  Li$_{3}$SiH$_{10}$ and Li$_{2}$SiH$_{6+\delta}$ which may be stabilised at high pressure. Our work provides pointers to future investigations on hydrogen-rich materials.
\end{abstract}

\maketitle

\section{\textbf{Introduction}}
Hypervalences are well established in chemistry, referring to aggregates, such as molecules, ions, hydrogen bonds, and other extended structures, in which main group atoms, as a result of their coordination with ligands, adopt a valence electron configuration that exceeds eight\cite{noury2002chemical}. The electronegativity of hydrogen is similar to that of $p$-block element central atoms, and this explains why hydrogen is rarely observed as ligand in hypervalent species\cite{treichel1967synthesis,seel1971umsetzung,beckers1993neue}. Linear bonding in hypervalent species is visually described by the three-center-four-electron ($3c-4e$) model \cite{doi:10.1021/ja01153a086,pimentel1951bonding,koutecky1974molecular} and it is generally acknowledged that a high polarity is essential to stabilise a hypervalent bond. The Lewis-Langmuir theory of valence attributes the stability of molecules to their ability to place their valence electrons, which appropriately paired off as chemical bonds into stable octets \cite{lewis1916atom,langmuir1919arrangement}. This theory can be said to have survived the advent of quantum mechanics, with the electron-pairs being replaced by doubly-occupied molecular orbitals or approximately-localized bond orbitals, and with the “cubical atom” being replaced by the directed valences of $p$-bonds, so that undistorted bonds form along perpendicular axes rather than towards the corners of tetrahedra. However, both types of theories have encountered considerable difficulties in treating the molecules formed by atoms of Groups V-VIIl of the periodic table in their higher valences, with the classical theory referring to “expanded octets” and the quantum mechanical theory invoking $d$-orbital participation in hybridization schemes involving large promotion energies\cite{musher1969hypervalente}. 

Some previous theoretical and experimental studies show that the alkali metal potassium(K) can form K-Si-H compounds, in which potassium silyl KSiH$_{3}$ exhibits a high hydrogen content of $4.1\;wt\%$ and maintains an excellently reversible reaction without any disproportionation through direct hydrogenation of the KSi Zintl phase near $130\;^{\circ}$C and ambient conditions, consistent with theoretical calculations \cite{janot2016catalyzed,chotard2011potassium,ring1961preparation,ring1961crystal,mundt1989metallderivate}. In $2012$, K$_{2}$SiH$_{6}$ with a cubic $Fm\overline3m$ structure was synthesized via reactions of $\text{K}_{2}\text{SiH}_{6} \rightleftharpoons 2\text{KH} + \text{Si} + 2\text{H}_{2}$ and $\text{K}_{2}\text{SiH}_{6} \rightleftharpoons \text{K} + \text{KSi} + 3\text{H}_{2}$ at pressures above $4$ GPa and temperatures between $450$ and $650$ $^{\circ}$C. This phase containes octahedral SiH$_{6}^{2-}$ and additional hypervalences of the central Si atoms \cite{puhakainen2012hypervalent}. We believe that there should be similar additional hypervalences for central Si atoms when combined with alkali metal hydrides at high pressure. 

In our work, alkali metal lithium (Li) was chosen to form Li-Si-H compounds. Here, the atomic mass of lithium is relative very light such that it enhances the hydrogen content to achieve the same stoichiometric ratio to metal as seen in K$_{2}$SiH$_{6}$ and similar compounds. In addition, we explore new ternary phases and investigate their peculiar structures and properties at high pressure, and find that these potential structures may be stable at ambient pressure.

\begin{figure*}[htbp]
    \centering
	\includegraphics[width=\textwidth]{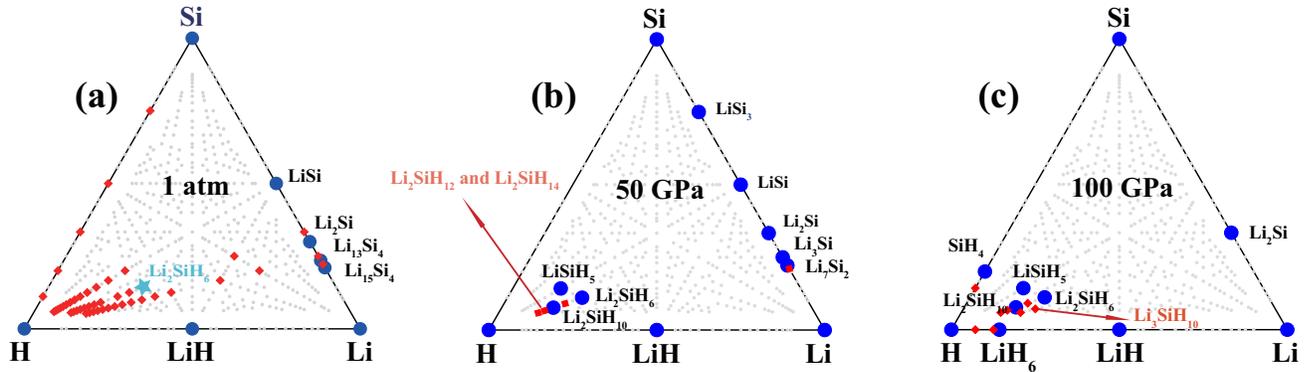}
	\caption{The ternary phase diagram of Li-Si-H at (a) 1 atm, (b) 50 GPa, and (c) 100 GPa. Details of phase diagrams are shown in (d). Big circles represent compounds which located on the convex hulls, and small red diamonds represent which didn't locate on the convex hulls. A metastable structure with	a ratio of Li$_{2}$SiH$_{6}$ is marked in pink star in the left plane below (a), which is dynamically stable .}
	\label{figure:CH}
\end{figure*}

\section{\textbf{Computational Details}}
We performed a full structure search of the ternary composition space bounded by Li-Si-H at $1$ atm, $50$ and $100$ GPa using the AIRSS (Ab Initio Random Structure Searching) method \cite{pickard2011ab,pickard2006high}, which enabled us to construct a ternary phase diagram in FIG \ref{figure:CH}. Full variable-composition predictions were firstly performed within $15000$ structures using random structure searches based on the AIRSS code at pressures of $1$ atm, $50$ and $100$ GPa. Then, fixed composition predictions were employed to further search structures for stable compounds by means of the AIRSS code as well. The CASTEP code\cite{segall2002first} was used for the AIRSS searches. The VASP (Vienna ab initio simulation packages) code\cite{kresse199614251} was used to optimize crystal structures and calculate the electronic properties, where the Perdew–Burke–Ernzerhof (PBE)\cite{perdew1996generalized} generalized gradient approximation (GGA)\cite{perdew1992pair} with the all-electron projector-augmented wave method (PAW)\cite{blochl1994projector} was performed. The electron-ion interaction was described by projector-augmented-wave potentials with the $1s^{1}$, $1s^{2}2s^{1}$ and $2s^{2}2p^{6}3s^{2}3p^{2}$ configurations treated as valence electrons for H, Li and Si, respectively. The kinetic cutoff energy of $800$ eV, and Monkhorst-Pack $\bm{k}$ meshes with grid spacing of $2\pi \times 0.03 \mathring{A}^{-1}$ were then adopted to ensure enthalpy convergence to less than $1$ meV/atom. Ground-state and semi-state structures for Li-Si-H compounds at $1$ atm, $50$ GPa and $100$ GPa were chosen in the calculations of the convex hulls. The structural parameters of these structures are listed in Supplemental Table S 2. The phonon calculations were performed in the PHONOPY code\cite{togo2008first}, for which the computational settings are described in the supplementary materials. To investigate the Si-H bonding characteristics, the ICOHPs were calculated as implemented in the LOBSTER\cite{deringer2011crystal} package, which provides an atom-specific measure of the bonding character of states in a given energy region.

\section{\textbf{Results and Discussions}}	
	
\begin{figure*}[htbp]
    \centering
   	\includegraphics[width=\textwidth]{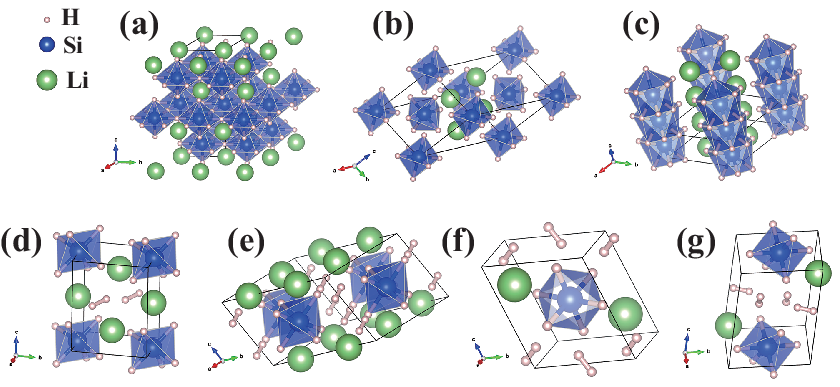}
	\caption{Crystal structures of Li-Si-H ternary compounds, which are (a) $R32\_$LiSiH$_{5}$ at 100 GPa, (b) $P2_{1}c\_$Li$_{2}$SiH$_{6}$ at 50 GPa, (c) $P\overline62m\_$Li$_{2}$SiH$_{6}$ at 1 atm, (d), (e), (f) and (g) are $P\overline1\_$Li$_{3}$SiH$_{10}$, $C2/m\_$Li$_{2}$SiH$_{10}$, $P\overline1\_$Li$_{2}$SiH$_{12}$ and $P\overline1\_$Li$_{2}$SiH$_{14}$ at 100 GPa.
	There are 1 type of SiH$_{5}^{-}$ in layers, and 2 types of SiH$_{6}^{2-}$, which are originally octahedral ions or tricapped trigonal prismatic ions.}
	\label{figure:Li-Si-H}
\end{figure*}

We have identified several new stoichiometries of Li-Si-H compounds, including LiSiH$_{5}$, Li$_{2}$SiH$_{6}$, Li$_{3}$SiH$_{10}$, and Li$_{2}$SiH$_{6+\delta}$, $\delta=4,6,8$. The structures of  these new Li-Si-H compounds are shown in FIG \ref{figure:Li-Si-H}. These results provide important pointers towards the searching for synthesis of new ternary hydrogen-rich metal hydrides. Furthermore, all of our new structures are hypervalent. There are no stable compounds located on the convex hull at $1$ atm. We find that LiSiH$_{5}$, Li$_{2}$SiH$_{6}$ and Li$_{2}$SiH$_{10}$ are stabilised by increased pressure and only appear on the convex hull in our calculations at $50$ and $100$ GPa. Li$_{3}$SiH$_{10}$, Li$_{2}$SiH$_{12}$, and Li$_{2}$SiH$_{14}$ are less than $2$ meV/atom above the convex hull at $50$ GPa or $100$ GPa, and thus are assumed to be potentially metastable. Phases located on the convex hulls at $50$ and $100$ GPa are Li$_{2}$SiH$_{6}$ and Li$_{2}$SiH$_{10}$ and our phonon calculations show that these are dynamically stable. By further calculation, some metastable phases like Li$_{3}$SiH$_{10}$, Li$_{2}$SiH$_{12}$, and Li$_{2}$SiH$_{14}$ are shown to be dynamically stable as well. 

In our further investigation we identify new structures formed that include Groups IV Si atoms. These include a dynamically stable layer-typed SiH$_{5}^{-}$, listed in FIG S 1. in the supplementary materials, which is the first such found as an inorganic structure. This is different from the SiH$_{5}^{-}$ previously identified in EtSiH$_{5}$ through the chemical reaction of Et$_{3}$SiH$_{2}^{\,-} + $SiH$_{4} \rightleftharpoons  $ SiH$_{5}^{-}$ + Et$_{3}$SiH in previous studies \cite{hajdasz1986hypervalent}. The only hypervalent all-hydride species reported before is the SiH$_{5}^{-}$ ion which was identified by mass spectrometry \cite{sheldon1984gas} as a product of the gas phase ion–molecule reaction. That form of SiH$_{5}^{-}$ ion has been reported stable with respect to the loss of H$^{-}$ but unstable with respect to decomposition into SiH$_{3}^{\,-}$ and H$_{2}$ \cite{taketsugu1995dynamic,pierrefixe2007hypervalence,pierrefixe2008hypervalent}. SiH$_{5}^{-}$ in EtSiH$_{5}$ is composed of one axial H-Si-H unit with $3$ H ligands perpendicular to the axial H-Si-H \cite{pierrefixe2007hypervalence}. We find unstable folded layer-typed SiH$_{5}^{-}$ ions in $P2_{1}c$-type LiSiH$_{5}$ (located on the convex hull at $50$ GPa) or $P2_{1}2_{1}2$-type LiSiH$_{5}$ (located on the convex hull at $100$ GPa), see FIG S 2. in supplementary material. We note that only flat layer-typed SiH$_{5}^{-}$ ions in the $R32$-type LiSiH$_{5}$ adopt a stable structure. The upper Si-H layer moves $\displaystyle \frac{2}{\sqrt{3}}d_{\text{Si-H}}$ from down layers alongside crystal orientation $[210]$ in $R32$-type LiSiH$_{5}$ such that $d_{\text{Si-H}}$ is the distance between central Si to H at $6c$ site (H on the Si-H plane), and the $R32$-type LiSiH$_{5}$ can be stable at the range of $60 \sim 150$ GPa, conforming to the $3c-4e$ model with $4$ H-Si-H units per central Si atom. One H-Si-H is axial and others are on the Si-H plane. We can assume that layer-type SiH$_{5}^{-}$ is composed of single EtSiH$_{5}$-type SiH$_{5}^{-}$ ions. We should note that the $3c-4e$ model does not explain why Si atoms can accommodate significant ligands in their valence shell, although it accounts for the bonding in hypervalent species.  
 
\begin{figure}[htbp]
    \centering
	\includegraphics[width=0.5\textwidth]{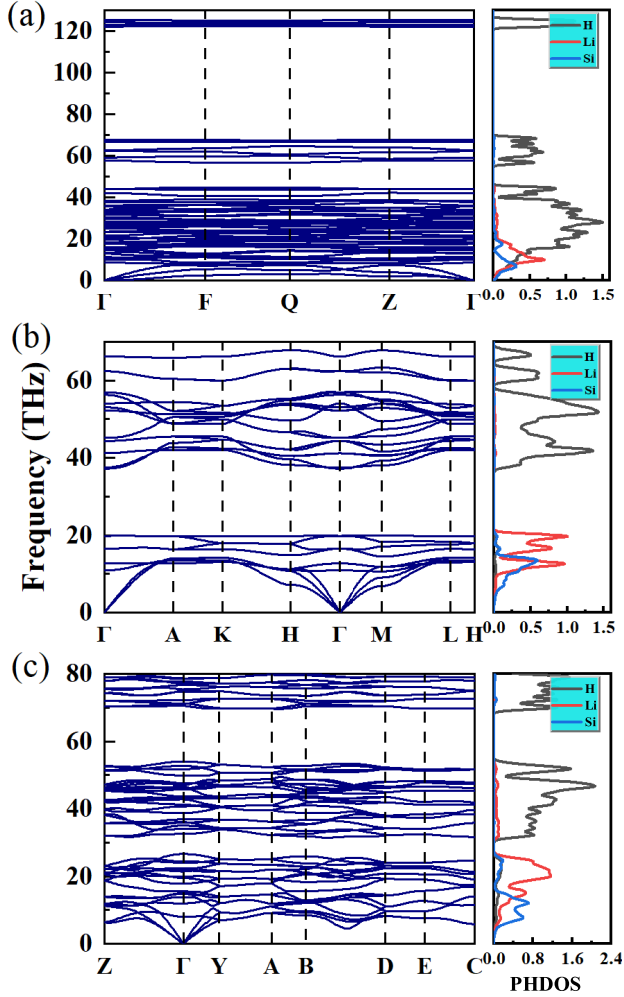}
	\caption{Calculated phonon dispersion relations (left panel) and densities of the phonon states (right panel) of (a) $R32\_$LiSiH$_5$, (b) $P2_{1}c\_\text{Li}_{2}\text{SiH}_{6}$, and (c) $C2/m\_$Li$_{2}$SiH$_{10}$ at 100 GPa.}
	\label{figure:phonon1}
\end{figure}
 	
\begin{figure}[htbp]
    \centering
	\includegraphics[width=0.5\textwidth]{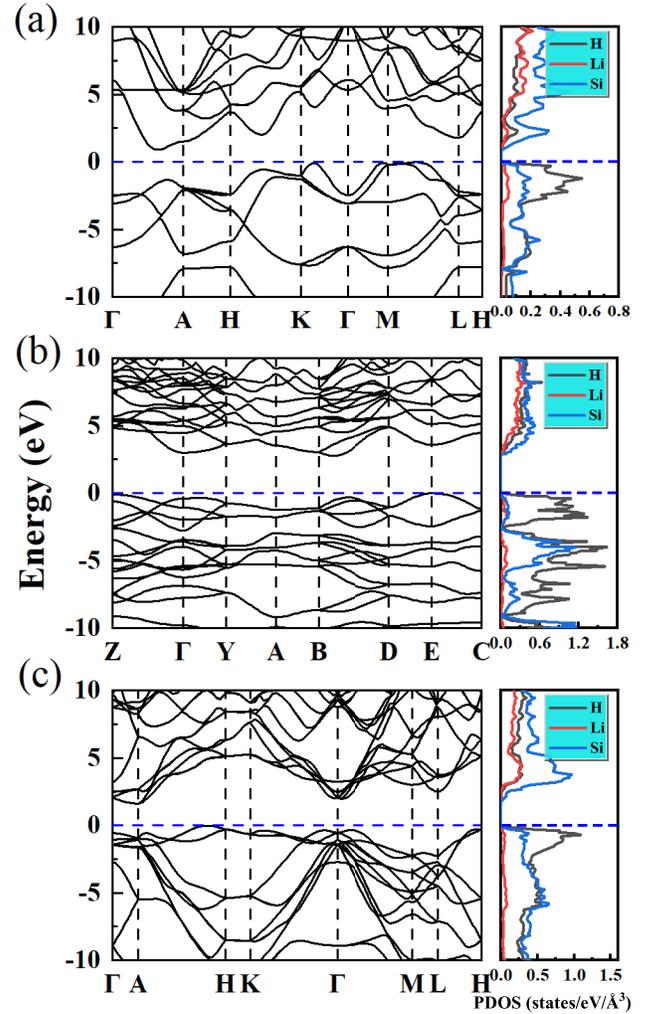}
	\caption{Calculated band structure (left panel) and projected density of states (right panel) of (a) $R32\_$LiSiH$_5$ at 100 GPa, (b) $P\overline62m\_\text{Li}_{2}\text{SiH}_{6}$ at 1 atm, and (c) $C2/m\_\text{Li}_{2}\text{SiH}_{10}$ at 10 GPa. The Fermi level is set to zero and depicted as the blue dash line.}
	\label{figure:BandD1}
\end{figure}
 	
We have, moreover, identified new tricapped trigonal prismatical SiH$_{6}^{2-}$ in the $P\overline62m$-type Li$_{2}$SiH$_{6}$ that does not follow the predictions of the $3c-4e$ model, whereas the original hypervalent octahedral structure SiH$_{6}^{2-}$ does obey the $3c-4e$ model where it occurs in the $P2_{1}c$-type Li$_{2}$SiH$_{6}$, Li$_{2}$SiH$_{6+\delta}$, and Li$_{3}$SiH$_{10}$. For compounds with stoichiometry Li$_{2}$SiH$_{6}$ we find three structures with space groups of $P\overline62m$, $P2_{1}c$, and $P\overline3$. $P2_{1}c$-type Li$_{2}$SiH$_{6}$ is located on the convex hull at $50$ GPa, but becomes dynamically unstable when decompressed to $20$ GPa. In contrast, $P\overline62m$-type Li$_{2}$SiH$_{6}$ is stable over the pressure range  $0\sim150$ GPa. The enthalpy calculation shows that the phase $P2_{1}c$ transforms to the phase $P\overline62m$ at $110$ GPa. Furthermore, $C2/m$-type Li$_{2}$SiH$_{10}$ is stable over the pressure range  $10\sim100$ GPa. Additional compounds that we identify include Li$_{3}$SiH$_{10}$, Li$_{2}$SiH$_{12}$, and Li$_{2}$SiH$_{14}$ which all adopt space group $P\overline1$. These all contain original hypervalent octahedral-structured SiH$_{6}^{2-}$ units with H$_{2}$ units. As expected, Li$_{x}$SiH$_{y}$ can be both thermodynamically and dynamically stable under high pressures. Li$_{3}$SiH$_{10}$ is stable at the pressure range of $70\sim100$ GPa. Li$_{2}$SiH$_{12}$ is stable at the pressure range of $50\sim100$ GPa. And Li$_{2}$SiH$_{14}$ is stable over the pressure range of $50\sim150$ GPa. For Li$_{3}$SiH$_{10}$ and Li$_{2}$SiH$_{6+\delta}$, a variety of bonds exist towards H$_{2}$ units and these are extended into different directions in their structures, which reduces the symmetries of these compounds to triclinic or monoclinic. At pressures greater than $200$ GPa, all ternary compounds become metastable and move off the convex hull, by more than $0.3$ eV/atom, and furthermore all of them are computed to be dynamically unstable. We examine the electronic properties of our new Li-Si-H compounds based on their equilibrium crystal structures. The Bader charge analysis\cite{bader1990international} was used to determine charge transfer, and the electron localization function (ELF)\cite{becke1990simple} was used to describe and visualize chemical bonds in molecules and solids. Bader charge analysis reveals charge transfer from Li/Si to H, as listed in Table S 1, suggesting that both Li and Si are electron donors and provide electrons to H atoms. For all the stable phases,  Li consistently loses approximate $0.8\sim0.9$e, and Li-ions form ionic bonds with Si-H anions. From FIG \ref{figure:BandD1} and FIG S 10 $\sim$ FIG S 15 in supplementary material, most of Li-Si-H compounds have large-band gaps, which shows that the majority of these Li-Si-H compounds are insulators.

  \begin{table*}[htbp]
   \caption{The ICOHPs for Si-H bonding in LiSiH$_{5}$, Li$_{2}\text{SiH}_{6}$, and Li$_{2}\text{SiH}_{10}$.}
       \begin{ruledtabular}
    \begin{tabular}{ccccc}  
		\textbf{Compounds} & \textbf{Pressure} & \textbf{Space Group} & \textbf{Bond length  ($\bm\mathring{A}$)} & \textbf{ICOHP (eV)} \\  
		\hline
		LiSiH$_{5}$ & $100$ GPa & $R32$ & $1.4584 \sim 1.6167$ & $-2.0772 \sim -1.9725$ \\
		Li$_{2}$SiH$_{6}$ & $1$ atm & $P\overline62m$ & $1.7510 \sim 1.7927$ & $-1.6513 \sim -0.2301$ \\
		Li$_{2}$SiH$_{10}$ & $10$ GPa & $C2/m$ & $1.5757 \sim 1.5853$ & $-0.9200 \sim -0.2982$ \\
    \end{tabular}
    \end{ruledtabular}
       \label{table:ICOHP} \\
  \end{table*}
  
We here propose a generally useful system to designate the bonding around the Si atoms. Such a bonding system is described as an N-X-L system\cite{perkins1980electrically}. N represents the number of total valence electrons, X is central atom, Si in our case, and L is the number of ligand atoms. There are $3$ different such systems which can thus be designated. The first is $10-\text{Si}-8$ $R32$-type LiSiH$_{5}$, which is thus different to the $10-\text{Si}-5$ seen in EtSiH$_{5}$ \cite{hajdasz1986hypervalent}. Additionally we identify $12-\text{Si}-9$ for $P\overline62m$-type Li$_{2}$SiH$_{6}$, and $12-\text{Si}-6$ for Li$_{2}$SiH$_{6+\delta}$ and $P\overline1\_$Li$_{3}$SiH$_{10}$ (Table S 3 in supplementary material). The bonding between H and Si atoms is covalent, composing hypervalent silicic anions SiH$_{5}^{-}$ or SiH$_{6}^{2-}$, which are strong and hard to disrupt at high pressure, as seen when we combine our data with the integrated crystalline orbital Hamiltonian population (ICOHP) results below \cite{dronskowski1993crystal}. For various ICOHP values, stronger Si-H bonds give larger negative values. Almost all of Li-Si-H compounds are semiconductors or insulators with energy gaps greater than $1.0$ eV, and even $P\overline1$-type Li$_{3}$SiH$_{10}$ is a poor  electrical conductivitor with high proportions of SiH$_{6}^{2-}$ and H$_{2}$ units. We have calculated the ICOHPs for Si-H bonds to characterise the orbital interaction in the hypervalent siliconhydride anions. All our results are presented in Table \ref{table:ICOHP}, and Table S 5 in the supplementary material. It is found that all the predicted compounds have high gravimetric hydrogen contents and volumetric hydrogen densities, suggesting that they may be potential hydrogen storage materials. $P\overline62m$-type Li$_{2}$SiH$_{6}$, at ambient pressure, has a very high volumetric hydrogen density of $175.04$ g/L with $12.51 wt\%$ theoretical gravimetric hydrogen content. Moreover, $P\overline1$-type Li$_{2}$SiH$_{14}$ has the highest volumetric hydrogen density of $352.31$ g/L, with $25.02$ $wt\%$ theoretical gravimetric hydrogen content.

\section{\textbf{Conclusions}}
In summary, we have explored the crystal structures and stabilities of compounds within the Li-Si-H ternary system under high pressure by employing \emph{ab initio} calculations as implemented in the AIRSS random structure search method. We predict the existences of several hydrogen-rich structures at low pressures, including $P\overline62m$-type Li$_{2}$SiH$_{6}$ at $1$ atm, and $C2/m$-type Li$_{2}$SiH$_{10}$ at $10$ GPa. Furthermore, we find highly hydrogen-rich structures in the form of $P\overline1$-type Li$_{2}$SiH$_{12}$ at $50$ GPa, $P\overline1$-type Li$_{3}$SiH$_{10}$ and $P\overline1$-type Li$_{2}$SiH$_{14}$ at $100$ GPa which accommodate high hydrogen content in the form of H$_{2}$ units inside. We find two new types of hypervalent ions. One of them is layer-typed SiH$_{5}^{-}$ in $R32$-type LiSiH$_{5}$ containing $4$ H-Si-H units per central Si atom, one is axial and others comprise the $2$-D layers. The other variety of hypervalent ion that we identify is tricapped trigonal prismatic SiH$_{6}^{2-}$ in $P\overline62m$-type Li$_{2}$SiH$_{6}$. This does not obey the $3c-4e$ model and dynamically stable at both ambient pressure and high pressure. The prediction of tricapped trigonal prismatic SiH$_{6}^{2-}$ may be a new challenge to the $3c-4e$ model. Based on our results we propose a new method for designing novel ternary hydrogen rich materials. We believe that our work provides useful guidance and waymarkers for future experimental synthesis of ternary alkali and alkaline earth metal hydrides.

\begin{acknowledgments}
This work was supported by National Natural Science Foundation of China (Nos. 11674122 and 11704143), Program for Changjiang Scholars and Innovative Research Team in University (No. IRT$_1$5R23).  X. Feng acknowledges China Scholarship Council. CJP acknowledges financial support from the Engineering and Physical Sciences Research Council [Grant EP/P022596/1] and a Royal Society Wolfson Research Merit award. Parts of calculations were performed in the High Performance Computing Center (HPCC) of Jilin University and TianHe-$1$(A) at the National Supercomputer Center in Tianjin.
\end{acknowledgments}


\end{document}